# Quasi-two-dimensional heterostructures (K$M_{1-x}$Te)(LaTe$_3$) ($M$ = Mn, Zn) with charge density waves

Jin-Ke Bao[1,2], Christos D. Malliakas[3], Chi Zhang[4], Songting Cai[4], Haijie Chen[1,3], Alexander J. E. Rettie[5], Brandon L. Fisher[6], Duck Young Chung[1], Vinayak P. Dravid[4], and Mercouri G. Kanatzidis[1,3,*]

[1] *Materials Science Division, Argonne National Laboratory, Lemont, Illinois 60439, United States,* [2]*Laboratory of Crystallography, University of Bayreuth, 95447 Bayreuth, Germany* [3]*Department of Chemistry, Northwestern University, Evanston, Illinois 60208, United States,* [4]*Department of Materials Science and Engineering, Northwestern University, Evanston, Illinois 60208, United States,* [5]*Electrochemical Innovation Lab, Department of Chemical Engineering, University College London, London WC1E 7JE, United Kingdom,* [6]*Center for Nanoscale Materials, Nanoscience and Technology Division, Argonne National Laboratory, Lemont, Illinois, 60439, United States.*

## Abstract

Layered heterostructure materials with two different functional building blocks can teach us about emergent physical properties and phenomena arising from interactions between the layers. We report the intergrowth compounds KLa$M_{1-x}$Te$_4$ ($M$ = Mn, Zn; $x \approx 0.35$) featuring two chemically distinct alternating layers [LaTe$_3$] and [K$M_{1-x}$Te]. Their crystal structures are incommensurate, determined by single X-ray diffraction for the Mn compound and transmission electron microscope (TEM) study for the Zn compound. KLaMn$_{1-x}$Te$_4$ crystallizes in the orthorhombic superspace group $Pmnm(01/2\gamma)s00$ with lattice parameters $a$ = 4.4815(3) Å, $b$ = 21.6649(16) Å and $c$ = 4.5220(3) Å. It exhibits charge density wave (CDW) order at room temperature with a modulation wave vector $\mathbf{q} = 1/2\mathbf{b}^* + 0.3478\mathbf{c}^*$ originating from electronic instability of Te-square nets in [LaTe$_3$] layers. The Mn analog exhibits a cluster spin glass behavior with spin freezing temperature $T_f \approx 5$ K attributed to disordered Mn vacancies and competing magnetic interactions in the [Mn$_{1-x}$Te] layers. The Zn analog also has charge density wave order at room temperature with a similar **q**-vector having the **c**\* component ~ 0.346 confirmed by selected-area electron diffraction (SAED). Electron transfer from [K$M_{1-x}$Te] to [LaTe$_3$] layers exists in KLa$M_{1-x}$Te$_4$, leading to an enhanced electronic specific heat coefficient. The resistivities of KLa$M_{1-x}$Te$_4$ ($M$ = Mn, Zn) exhibit metallic behavior at high temperatures and an upturn at low temperatures, suggesting partial localization of carriers in the [LaTe$_3$] layers with some degree of disorder associated with the $M$ atom vacancies in the [$M_{1-x}$Te] layers.

# Introduction

Hetero-layered structures, consisting of two or more chemically different types of layers stacked in an alternating sequence, can exhibit emergent physical properties. For example, the discovery of cuprate[1] and Fe-based[2] high-$T_c$ superconductors containing $[CuO_2]^{2-}$ and $[FeAs]^-$ layers, respectively, have been in fact hetero-layered structures. Such materials harbor special states such as charge density waves (CDW) or spin density waves (SDW) which can be manipulated via external chemical or physical stimuli.[3-4] It is natural to wonder what properties may appear when combining distinct functional layers in a single compound and how they may couple to one another. Hetero-layered structures can provide a versatile testbed to explore antagonistic interactions such as between superconductivity and magnetism that may be present in different layers, exemplified by the novel magnetic superconductor $RbEuFe_4As_4$ with both superconducting [FeAs] layers and ferromagnetic Eu layers.[5-6] In addition, the $Sr_2M'_3As_2O_2$ ($M'$ = Cr, Mn) present a typical example to successfully intergrow $[M'O_2]^{2-}$ and $[M'As]^-$ along $c$ axis with $Sr^{2+}$ ions acting as a separating layer to connect the stacked $[M'O_2]^{2-}$ and $[M'As]^-$ layers and balance the charge in those compounds.[7-8] They exhibit complicated magnetism and magnetic interaction between $[M'O_2]^{2-}$ and $[M'As]^-$ layers.[7-8] A distinct type in this class are the so-called misfit layer compounds $([RQ]_{1+\delta})_m(M''Q_2)_n$ ($R$ = Pb, Sn, Bi, rare earth elements; $M''$ = Ti, V, Cr, Nb, Ta; $Q$ = S, Se),[9-10] where each repeat unit consists of $m$ layers of rock-salt [$RQ$] and $n$ layers of transition metal dichalcogenide [$M''Q_2$] blocks. This stacking forms an incommensurate modulated heterostructure due to the mismatch of the lattice parameter from two sublattices along one axis. Since electron transfer between different functional layers can happen because of the different chemical potentials,[11-12] we can expect a large breadth of chemical characteristics, structure types and band structures from the hetero-layered family of solids.

Most hetero-layered structures have been discovered by accident given that a substantial knowledge gap exists on how to reliably synthesize such structures. There are a number of strategies one can envision about creating hetero-layered structures.[13-14] Here we show a direct synthesis approach by solid state reactions of appropriate precursors. One important question is what are the defining chemical characteristics and structural attributes that must be present in the reaction or selection of building blocks for a successful synthesis. How much can the two layered sublattices differ before the hetero-layered compounds can be destabilized? What is the role of "chemical compatibility" between the two slabs to favor the hetero-layered motif, as opposed to a phase separate mixture of homo-layered or other phases? To address such questions will require development of suitable synthesis science for hetero-layered structures and the results reported here are a contribution in this direction.

Following the general idea of creating new functional heterostructures, see Figure 1, we tried intergrowing a functional building block Te-square nets hosting a CDW order[15] and anti-PbO-type building blocks.[4] Our synthesis yielded two Q2D heterostructures $KLaM_{1-x}Te_4$ ($M$ = Mn, Zn) with intact intergrown $[LaTe_3]$ and $[KM_{1-x}Te]$ layers. Single crystal X-ray diffraction analysis, transmission electron microscopy (TEM), coupled with physical property measurements including magnetic susceptibility, charge transport and heat capacity indicate that these are CDW compounds with substantial electron transfer from $[KM_{1-x}Te]$ to $[LaTe_3]$ layers. The Mn analog features massive Mn vacancies that give rise to cluster spin glass (SG) behavior with a spin freezing temperature $T_f \approx 5$ K.

## Experimental Section

**Synthesis.** Potassium metal pieces (99.5%, Sigma), lanthanum powder and pieces (99.7%, Alfa Aesar), manganese powder (99.95%, Alfa Aesar), zinc powder (99.9%, Alfa Aesar) and tellurium chunks (99.999%, Plasmaterials) were used as the reagents for the synthesis as received. The lanthanum powder was sealed in an ampoule under argon when it was delivered and the ampoule was opened inside the glovebox. All the reactions were performed in evacuated fused-silica ampoules (I.D. = 13 mm) under a vacuum of ~ $3 \times 10^{-3}$ mbar. We prepared two binary precursors, $LaTe_3$ and $K_2Te_3$ for synthesis of $KLaM_{1-x}Te_4$. La and Te chunks were used to prepare the precursor $LaTe_3$. Fine La powder should be avoided in this reaction due to the violent exothermic reaction. Typically, 0.47 g (3.4 mmol) La pieces and 1.31 g (10.2 mmol) Te pieces were loaded into the ampules, which were heated up to 1073 K in 24 h and held at this temperature for 24 h. $LaTe_3$ was ground into fine powder before it was used for further reactions. $K_2Te_3$ was synthesized by reacting K and Te pieces in a stoichiometric composition. 0.31 g (8 mmol) of K pieces were put on the top of 1.53 g (12 mmol) of Te pieces in a fused-silica ampule. The mixture was slowly heated to 393 K in 20 h and held for 12 h before it cooled down to room temperature. Splashes of K metals were often observed from the violent reaction of K metal with Te pieces. The reaction was then rapidly heated to 723 K in 3 h and held for 2 h before the furnace was shut off. A single black ingot of $K_2Te_3$ was easily removed from the ampule. The obtained $K_2Te_3$ was ground and used for $KLaMn_{1-x}Te_4$ crystal growth.[16-17]

$K_2Te_3$, Mn or Zn powder, Te pieces and La powder were mixed in a composition "$K_8LaMn(Zn)_{0.65}Te_{20}$", which were loaded into a fused-silica ampule. The mixture was heated to 973 K with a rate of 15 K h$^{-1}$, held for 24 h and then cooled down to 573 K at a rate of 2 K h$^{-1}$. The $KTe_y$ flux was then removed using *N,N*-Dimethylformamide under N$_2$ gas flow to obtain shiny brown square-plate crystals, see the insets of Figure 2. Polycrystalline $KLaMn_{1-x}Te_4$ (total mass ~ 1 g) was synthesized by directly loading Te pieces, LaTe$_3$, Mn powder and K pieces into a fused-silica ampule before being flame-sealed under vacuum in an appropriate composition. The mixture was slowly heated to 673 K in 20 h and held for 24 h. The pre-reacted mixture was ground in an agate mortar and heated to 773 K for another 48 h. Finally, the KLaMn$_{1-x}$Te$_4$ powder was ground again, pressed into a pellet and annealed at 623 K for another 48 h. This procedure gave the best phase-pure KLaMn$_{0.65}$Te$_4$ product judged by powder X-ray diffraction (PXRD) patterns, see Figure S1 in supporting information (SI). Polycrystalline sample KLaZn$_{1-x}$Te$_4$ was synthesized using the same procedure under the slightly different stoichiometry KLaZn$_{0.67}$Te$_4$ to get the purest phase. KLa$M_{1-x}$Te$_4$ ($M$ = Mn, Zn) are thermodynamically stable phases. The formation of these phases does not closely depend on the initial starting materials as long as the composition is close to KLa$M_{1-x}$Te$_4$. The synthetic methods described here lead to samples in the best quality.

**Powder X-ray Diffraction**. PXRD were performed by a PANalytical diffractometer X'Pert with a Cu K$_\alpha$ radiation. KLa$M_{1-x}$Te$_4$ ($M$ = Mn, Zn) samples are slightly air-sensitive and deteriorate into other phases such as Te element when exposed in air for a long time, see the case of Zn compound in Figure S2 in SI. Thus a thin Kapton film was covered on the plate sample holder to protect samples from being exposed to air. KLaMn$_{1-x}$Te$_4$ and KLaZn$_{1-x}$Te$_4$ have the same PXRD pattern with only a slight shift in the diffraction angle, indicating they have the same average crystal structure with different lattice parameters. Plate-like crystals were oriented on the sample stage to observe only (0$k$0), see Figure 2. The lattice parameter $b$ was extracted by least squares fitting of selected high-intensity Bragg peaks after introducing a zero-shift parameter.

**Single X-ray Diffraction.** Single crystal diffraction data of KLaMn$_{1-x}$Te$_4$ were collected on the diffractometer STOE IPDS 2T equipped with a 34-cm image plate detector at 293 K. Since KLaMn$_{1-x}$Te$_4$ crystals are air-sensitive, crystals were sealed in glass capillaries before the single

X-ray diffraction measurements. Basic lattice parameters and the modulation vector were extracted from the instructed reciprocal space from all the measured frames by X-area software package.[18] Data reduction of main Bragg and satellites reflections and absorption correction were also done by the X-area package. The modulated structure of KLaMn$_{1-x}$Te$_4$ was solved and refined by the JANA2006[19] software using the (3+1)-dimensional superspace method. Two modulation waves consisting of positional displacement and thermal atomic displacement parameters (ADP) were applied to each atom. The basic refinement results are summarized in Table 1 in the main text. Detailed structure information including atomic positions in each unit cell, anisotropic displacements, two modulation waves, bond lengths can be found in Table S1-5 in SI. The crystals of KLaZn$_{1-x}$Te$_4$ were very soft and malleable which makes it hard to obtain good single crystal diffraction data for structural refinements. The dense dislocation defects observed by TEM in the KLaZn$_{1-x}$Te may also be attributed to this crystal behavior, see Figure S7 in SI.

**Specific Heat**. Specific heat measurements were done on the commercial Physical Property Measurement System (DynaCool, Quantum Design) using a thermal relaxation method. A tiny amount of N grease was applied to the heat capacity puck and the addenda data of the puck and N grease was collected first. Polycrystalline samples KLaMn$_{1-x}$Te$_4$ or KLaZn$_{1-x}$Te$_4$ were then cut and polished into a suitable dimension (~ 2 × 2 × 0.3 mm$^3$) for specific heat measurements. Two measurements of specific heat in the same batch were done for both KLaMn$_{1-x}$Te$_4$ and KLaZn$_{1-x}$Te$_4$. The two data sets are almost same below 100 K for both compounds while they have a small difference at high temperature especially for the case of KLaZn$_{1-x}$Te$_4$. This is probably due to N grease or impurity phases of which the influences to the two separately prepared samples are slightly different, see Figure S4 in SI. The data below 60 K with no influence from those two possible factors are presented and analyzed in the main text.

**Magnetic Susceptibility**. DC magnetic susceptibility measurements were performed on the commercial Magnetic Property Measurement System (MPMS3, Quantum Design) in a vibrating sample magnetometer mode. Lots of KLaMn$_{1-x}$Te$_4$ single crystals were carefully selected from the same batch. They were mounted on the quartz sample holder by N grease, where the crystals (~ 3 mg) were oriented and a magnetic field was applied parallel to *ac* plane. A small piece of KLaMn$_{1-x}$Te$_4$ or KLaZn$_{1-x}$Te$_4$ cut from a polycrystalline pellet was mounted on a brass holder for magnetic property measurements. Zero field cooling (ZFC) and field cooling (FC) procedures were done to check any magnetic hysteresis in the sample. Single crystals and polycrystalline pellets of KLaMn$_{1-x}$Te$_4$ gave similar temperature-dependent magnetic susceptibilities with exactly the same temperature of the cusp $T_f$ under ZFC procedures. AC magnetic susceptibility measurements were carried out on the Physical Property Measurement System with frequencies in the range from 11 to 9995 Hz. Several pieces of polycrystalline ingots (~ 500 mg) were loaded into a liquid sample holder and an AC driving field of 10 Oe was applied. The freezing temperature $T_f$ at each frequency was extracted by fitting the peak curve with a polynomial function and then finding the local maximum near the cusp position.

## Results and Discussion

**New Hetero-layered Compound Exploration.** Following the basic idea of intergrowth of two functional layers, see Figure 1, we initially targeted hetero-layered compounds such as those combining [LaTe$_3$] (CDW layer) and [KFeAs] (high-$T_c$ superconducting layer). These attempts were unsuccessful presumably because of significant in-plane lattice mismatch between LaTe$_3$ ($a$ ≈ 4.3-4.4 Å )[15] and *A*FeAs (*A* = K, Na) ($a$ ≈ 3.9-3.95 Å).[21] We then tried KMnAs (magnetic layer) with an in-plane lattice parameter ($a$ ≈ 4.39 Å)[22] close to that of LaTe$_3$, however, no intergrown products could be detected by PXRD suggesting that transition metal arsenide layers may be incompatible with Te-square lattices. KMnAs was transformed into a ternary compound telluride KMnTe$_2$ when it reacted with LaTe$_3$ at high temperatures. A switch to transition metal tellurides was then adopted which led to two hetero-layered compounds KLa$M_{1-x}$Te$_4$ ($M$ = Zn, Mn) with Te-square nets and [$M_{1-x}$Te] layers. We also tried other transition metals (Fe, Co, Ni, Ru, Rh, Pd, Re, Ir, Pt) to replace Mn or Zn. None of them could form this intergrowth structure. The reason may be the lack of stable [$T_{1-x}$Te] ($T$ = transition metal) blocks with these metals or significant in-plane lattice mismatch between [KFe$_{1-x}$Te] ([FeTe] blocks, $a \approx 3.8$ Å)[23] and [LaTe$_3$] ($a \approx 4.3$-4.4 Å).[15] As shown in Figure 1, two functional layers can carry electric charges, which are charge-balanced and bonded together by extra layers between them. As a result, in-plane lattice parameter match plays an important role in stabilizing this kind of intergrowth structure as discussed below.[14] However, in-plane lattice parameter match is not a requisite for many misfit layer compounds where the in-plane lattice parameters from two the different layers along do not completely match.[9] The lattice mismatch drives them to form an incommensurate modulated structure along one axis while they share the same periodicity along the other two axes. Two favorable factors stabilize such misfit layer structures: a) Strong intralayer but weak interlayer bonding; b) charge transfer between those two layers which acts an essential glue to stick them together by the attractive Coulomb interaction from oppositely charged layers.[24-27] Charge transfer from [K$M_{1-x}$Te] to [LaTe$_3$] in KLa$M_{1-x}$Te$_4$, which will be proved below, likely helps stabilize the intergrowth structure.

**Crystal Structure.** The basic crystal structure of KLaMn$_{1-x}$Te$_4$ can be interpreted as the intergrowth of alternating [LaTe$_3$][15] and [KMn$_{1-x}$Te][28] layers, see Figure 3. Each [LaTe$_3$] layer consists of a buckled rock-salt [LaTe] layer sandwiched by two distorted Te-square nets. The [KMn$_{1-x}$Te] layers are composed of anti-PbO-type [Mn$_{1-x}$Te] layers (with Mn vacancies) and charge balancing K ions. When $x$ is 0.5, the layers become identical to those in KMnTe$_2$.[28] In the [Mn$_{1-x}$Te] layers, the MnTe$_4$ tetrahedra are connected to each other by sharing edges and corners while they only share corners in KMnTe$_2$ ($x$ = 0.5).[28] The average structure of KLaMn$_{1-x}$Te$_4$ at room temperature was solved and refined from single crystal X-ray diffraction data in the orthorhombic space group *Pmnm* with lattice parameters $a$ = 4.4815(3) Å, $b$ = 21.6649(16) Å and $c$ = 4.5220(3) Å. The $b$ parameter is the stacking axis and directed perpendicular to the crystal plate, Figure 2(a). The KLaZn$_{1-x}$Te$_4$ is isostructural with a smaller $b$ lattice parameter (21.46 Å), see Figure 2(b). The corresponding parameter in KLaMn$_{1-x}$Te$_4$ is very close to the summation of the thickness of [LaTe$_3$] ($b/2$ = 13.113 Å) and [KMn$_{0.5}$Te] ($c/2$ = 7.454 Å) layers, providing a straightforward criterion to check the success of the intergrowth during the exploration using PXRD data. The in-plane lattice parameters in LaTe$_3$ ($a$ = 4.3944 Å, $c$ = 4.4076 Å)[15] are comparable to KMnTe$_2$ ($a$ = $b$ = 4.5110),[28] which is important to stabilize the heterostructure KLaMn$_{1-x}$Te$_4$ as exemplified in many intergrowth compounds.[29-30] The average distance between K and Te within [KMn$_{1-x}$Te] layers is around 3.6 Å, only slightly smaller than the value (3.7 Å) between [KMn$_{1-x}$Te] and [LaTe$_3$] layers, indicating a significant interlayer bonding. The average bond length between Mn and Te in the [Mn$_{1-x}$Te] layer is 2.772 Å, which is slightly larger (2.757 Å) than that observed in the [Mn$_{0.5}$Te] layer of another hetero-layered compound BaFMn$_{0.5}$Te.[20] Although the value of Mn vacancy is close to a rational number 1/3, the diffraction data don't support the scenario of Mn-vacancy ordering in the [Mn$_{1-x}$Te] layer. The spin-glass behavior at low temperature shown below in KLaMn$_{1-x}$Te$_4$ is also consistent with the disordered model of Mn vacancies in the [Mn$_{1-x}$Te] layer, different from the case in BaFMn$_{0.5}$Te which enters an antiferromagnetic ground state because of the Mn vacancy ordering in the [Mn$_{0.5}$Te] layer.[20] The refined Mn vacancy from a disordered model is $x \approx 0.35$ which is close to both the composition used in synthesis of KLaMn$_{1-x}$Te$_4$ and the composition determined by EDS in SI. This agreement on the $x$ in the compositions above also applies to the Zn case. Since the actual value of Mn or Zn vacancies does not influence the conclusions, for convenience the composition KLa$M_{1-x}$Te$_4$ ($M$ = Mn, Zn) with $x \approx 0.35$ is adopted in the whole article, emphasizing that massive vacancies exist in these compounds.

The square-Te nets in KLaMn$_{1-x}$Te$_4$ are introduced by the [LaTe$_3$] layers whose CDW order has been well confirmed and studied in the $RE$Te$_3$ ($RE$ = rare earth elements) compounds.[15, 31-32]. As expected, KLaMn$_{1-x}$Te$_4$ also exhibits CDW order originating from the same square-Te nets at room temperature. The actual transition temperature can be very high, which is not determined in this study and needs to be further investigated by high-temperature physical property measurements. This CDW order imposes an orthorhombic distortion of the $ac$ plane in KLaMn$_{1-x}$Te$_4$, which gives an incommensurate modulation wave vector $\mathbf{q} = 1/2\mathbf{b}^* + 0.3478\mathbf{c}^*$. This $\mathbf{q}$-vector is similar to $1/2\mathbf{b}^* + 0.3861\mathbf{c}^*$ in KLaCuTe$_4$ having a CDW order above room temperature but it is significantly different from the value $0.2751\mathbf{c}^*$ in LaTe$_3$.[15, 17] The magnitude of $\mathbf{q}$-vector in the square-Te nets is strongly dependent on the electronic filling in the CDW. Assuming the $\mathbf{q}$-vector in the homo-layered LaTe$_3$ represents a baseline where no electronic transfer exists from another layer, the interlayer coupling between two different layers of KLaMn$_{1-x}$Te in the form of electron transfer between the heterolayers can be assessed by the different $\mathbf{q}$-vector.

The modulated structure of KLaMn$_{1-x}$Te$_4$ adopts the superspace group $Pmnm$(01/2$\gamma$)$s$00, see Table 1. The atomic displacement amplitudes of the modulated waves in Te-square nets (Te(2) and Te(3) atoms) are much larger than other atoms, see Table S3 in SI, proving that CDW order is originating from the electronic instability of those nets. The CDW in the Te-square nets in KLaMn$_{1-x}$Te$_4$ causes distortions from the ideal square-net motif with Te(2)-Te(3) distances ranging from 2.940(7) to 3.360(7) Å, see lower panel of Figure 4. The distorted Te-square nets mainly consist of Te trimers separated by monomers and dimers with a Te-Te distance threshold of 3.08 Å, see the upper panel of Figure 4. The distribution of those Te fragments in KLaMn$_{1-x}$Te$_4$ is different from KLaCuTe$_4$[12] and LaTe$_3$[15], suggesting different degrees of electronic density on the [LaTe$_3$] layer. Although the detailed modulation structure of KLaZn$_{1-x}$Te$_4$ has not been resolved, it is likely similar to the Mn analog as shown in the SAED pattern below. The formula can be written as [K$^+$M$^{2+}_{1-x}$Te$^{2-}$]$^{1-2x}$[Te$_2^{2x-2}$La$^{3+}$Te$^{2-}$]$^{2x-1}$ ($x \approx 0.35$). This scenario is also supported by an enhanced electronic specific heat coefficient γ compared to the compound LaTe$_3$ shown below.

KLaZn$_{1-x}$Te$_4$ adopts the same average structure as KLaMn$_{1-x}$Te$_4$, which is supported by the PXRD patterns, see Figure S1. However, proving the existence of the modulation structure from single X-ray diffraction was challenging because of the low-quality data from the as-grown crystals. Therefore, we performed a scanning/transmission electron microscopy (S/TEM) investigation of its microstructure. Figure S7(a) is a typical high angle annular dark field (HAADF) image of the specimen, where a huge amount of dislocation arrays with brighter contrast can be observed; these dense dislocations can be induced by the intrinsic Zn vacancies in the sample. The existence of such high defect level will affect mechanical properties of the single crystal, making it hard to perform large area structural analyses on it.

Figure 5(a) shows the selected area electron diffraction patterns (SAED), where the main diffraction spots can be well indexed to the average orthorhombic structure (space group *Pmnm*) along the [010] zone axis. Moreover, extra spots along **c*** are showing up as highlighted with a yellow arrow. These spots match the second order satellites spots, indicating the existence of modulation structure with a **q**-vector having the **c*** component ~0.346 similar to KLaMn$_{1-x}$Te$_4$. These results are consistent with the single-crystal X-ray diffraction of KLaMn$_{1-x}$Te$_4$ sample.

Figure 5(b) displays the corresponding high-resolution TEM image of KLaZn$_{1-x}$Te$_4$. The measured interplanar $d$-spacings of (002), (200) and ($\bar{1}$01) planes in HRTEM are 2.31, 2.26, and 3.21 Å, respectively, in good agreement with the refinement results from PXRD.

**Charge Transport Properties.** Both KLaMn$_{1-x}$Te$_4$ and KLaZn$_{1-x}$Te$_4$ exhibit metallic behavior with moderately high resistivity near room temperature, see Figure 6(a). The resistivity of KLaZn$_{1-x}$Te$_4$ saturates to ~ 26 mΩ cm at 300 K whereas that of KLaMn$_{1-x}$Te$_4$ increases significantly below ~ 50 K, indicating a semiconducting-like behavior at low temperature. The resistivity upturn does not obey an exponential growth with decreasing the temperature, see Figure S8. The resistivity values at room temperature for both KLa$M_{1-x}$Te$_4$ ($M$ = Mn, Zn) are much larger than KLaCuTe$_4$ where there are no Cu vacancies in the [KCuTe] layer.[16-17] The larger values and low temperature upturn in resistivity are probably related to the massive random vacancies in the [K$M_{1-x}$Te] layer acting as scattering centers. Hall resistivity data indicates that hole carriers dominate over the whole temperature range in KLaMn$_{1-x}$Te$_4$ and the carrier concentration with a magnitude of $10^{20}$ cm$^{-3}$ slightly decreases with decreasing temperature, see Figure 6(b). The hole carrier concentration at 2 K (1.1 × $10^{20}$ cm$^{-3}$) is much smaller than the value (7.0 × $10^{20}$ cm$^{-3}$) in LaTe$_3$ at the same temperature[33] further supporting the scenario that electrons transfer from the [KMn$_{1-x}$Te] block to the Te square nets in the [LaTe$_3$] block.

The metallic character in KLa$M_{1-x}$Te$_4$ stems mainly from the 2D Te-square nets of the structure, which is known to be a metal despite the presence of the largely gapped CDW associated with the layer that contains the modulated square net.[31-32] The [$M_{1-x}$Te] component acts as an insulating layer as proved in the insulating behavior in BaFMn$_{0.5}$Te with almost identical [Mn$_{1-x}$Te] layer and will show unusual magnetic properties due to the local magnetic moment of Mn atoms.[20] The absence of obvious resistivity anomaly at the temperature where a spin glass freezing transition from magnetic Mn atoms occurs at ~ 5 K (see magnetic properties below) for KLaMn$_{1-x}$Te$_4$ is consistent with the spatial separation and greater 2D character of the conduction electrons and localized spins in the structure. The non-zero electronic specific heat coefficients obtained from the heat capacity data, presented below for these two compounds, are consistent with a non-vanishing density of states at Fermi level as expected for metallic systems. The low temperature upturn in resistivity for the two compounds indicates that they are on the boundary of metal-insulator-transition possibly due to the Anderson localization[34-35] originating from the disorder in these Q2D systems, just as the case in LaTe$_{1.95}$ with certain amount of Te vacancies in the Te square nets[36] while LaTe$_3$ without any intrinsic disorder shows metallic behavior all the way down to 2 K.[31]

**Magnetic Properties**

**DC Magnetic Susceptibility.** Temperature-dependent DC magnetic susceptibilities under ZFC and FC conditions with a magnetic field $H$ = 1000 Oe applied in the *ac* plane for KLaMn$_{1-x}$Te$_4$ are shown in Figure 7(a). A significant bifurcation of ZFC and FC in the curve appears at ~ 5 K where there is a cusp feature in the ZFC curve. The cusp and bifurcation move to lower temperatures and the sharpness of the cusp is smoothed out when the applied magnetic fields increase, see Figure 7(b). The cusp in the ZFC curve is due to the SG freezing behavior which will be confirmed by the AC magnetic susceptibility below. The cusp point is called the spin freezing temperature $T_f$.[37] Field-dependent magnetization at 150 and 300 K shows a linear behavior from 0 to 7 T while it exhibits a non-linear behavior at 2 and 10 K, see Figure 7(c). The field-dependent magnetization at 2 K also shows small hysteresis at low magnetic fields with a coercive field of ~ 800 Oe, see the inset of Figure 7(c), consistent with the spin freezing scenario below $T_f$.

The magnetic susceptibility data in KLaMn$_{1-x}$Te$_4$ above 150 K can be well fit by a modified Curie-Weiss formula (the inset of Figure 7(a)): $\chi_0 + \frac{C}{T-\theta}$, where $\chi_0$ is a constant value contributed by the temperature-independent diamagnetic part from electron orbital magnetism and the Pauli paramagnetic part from itinerant electrons in the system; $C$ is a Curie constant from which effective magnetic moment for each unit can be calculated; $\theta$ is the Curie-Weiss temperature whose value indicates the effective magnetic interaction strength between those magnetic moments.[38-39] The Curie-Weiss temperature $\theta$ obtained from the fitting is large and negative at −234 K, indicating strong antiferromagnetic interactions between Mn ions in the Mn-substructure at high temperatures but to a smaller degree than in BaFMn$_{0.5}$Te (−557 K). The effective magnetic moment for each Mn atom is 2.58 $\mu_B$, and is significantly smaller than the high spin state (5.91 $\mu_B$) for Mn$^{2+}$. Magnetic Mn ions in a tetrahedron [MnQ$_4$] ($Q$ = S, Se, Te) usually show a high spin state as exemplified by many examples such as K$_6$MnQ$_4$ ($Q$ = S, Se, Te) (~ 5.9 $\mu_B$)[40], BaFMn$_{0.5}$Te (~ 6.0 $\mu_B$)[20] and $A$MnTe$_2$ ($A$ = K, Rb, Cs) (~ 4.3-4.9 $\mu_B$).[28, 41-42] The smaller magnetic moment in KLaMn$_{1-x}$Te$_4$ may be due to the significant hybridization of Mn-3$d$ and its ligand Te-5$p$ electrons as the case in the BaMn$_2$$Pn$$_2$ ($Pn$ = P, As, Sb, Bi)[43-46] and $A$$_2$Mn$_3$Se$_4$ ($A$ = Rb, Cs).[47] The [KMn$_{1-x}$Te] layer in KLaMn$_{1-x}$Te$_4$ is basically an insulating layer with localized Mn magnetic moments. It is intercalated between two metallic Te-square nets, which may increase the degree of itinerancy of 3$d$ electrons of Mn ions in the [KMn$_{1-x}$Te] insulating layer due to the proximity effect. Such a phenomenon has also been observed in the compound Sr$_2$Cr$_3$As$_3$O$_2$[8] where the magnetic moment of the localized nature of Cr in the CrO$_2$ plane, sandwiched by two metallic Cr$_2$As$_2$ layers, is significantly reduced. Another scenario may be electron transfer between the Mn containing layer and the Te containing layer creating a small change in oxidation state in certain Mn atoms which reduces the overall average



effective moment. Such an electron transfer is implied by the heat capacity data below. The physical origin of the reduced magnetic effective moment in KLaMn$_{1-x}$Te$_4$ needs to be uncovered by further detailed investigations, which are beyond the study of this work.

When the magnetic field reaches 7 T the magnetization of KLaMn$_{1-x}$Te$_4$ at 2 K is only 0.12 $\mu_B$/Mn, much smaller than magnetic moment of Mn ions (~ 2 $\mu_B$/Mn) from the Curie-Weiss fitting mentioned above and points to antiferromagnetic interactions dominating between the Mn ions. The dominant antiferromagnetic interaction is also inferred from the field-dependent cusp temperature behavior shown above. A small bifurcation of the ZFC and FC curves appears at ~ 90 K, see Figure 7(a), which is also observed in our purest polycrystalline samples (Figure 7(d)). However, there is no anomaly at ~ 90 K in the magnetic susceptibility seen in less pure polycrystalline samples, see Figure S6 in SI. As a result, the anomaly is not intrinsic and probably arises from the magnetic transition from a tiny amount of MnTe$_2$, whose magnetic transition happens to be at ~ 90 K.[48] The magnetic susceptibility data of the KLaZn$_{1-x}$Te$_4$ samples are essentially temperature-independent at high temperatures with an upturn from paramagnetic impurities at low temperatures, consistent with a metallic Pauli paramagnetic behavior and non-magnetic Zn$^{2+}$ in this compound.

**AC Magnetic Susceptibility.** In order to prove the SG behavior in KLaMn$_{1-x}$Te$_4$ and further study its dynamic properties, frequency-dependent AC magnetic susceptibility measurements were used, see Figure 8. The real components of AC magnetic susceptibilities $\chi'$, measured under different frequencies, have similar behavior to the DC magnetic susceptibility while the imaginary components $\chi''$, a quantity representing energy dissipation, start to grow rapidly at around the cusp temperature and drop again at lower temperatures, see Figure 8(a) and (b). The cusp temperature in $\chi'$ moves to higher values with increasing applied AC magnetic field frequency and below the



cusp temperature the value of χ′ decreases, which are two typical characteristics of SG systems.[37] This cusp is called the spin freezing temperature $T_f$, see Figure 8(a). The edge where the value of χ″ increases rapidly also moves to higher temperature with increasing frequency, consistent with the case in χ′, see Figure 8(b). When a static magnetic $H$ = 500 Oe is applied, the cusp temperature in χ′ remains almost unchanged but it becomes broader, see Figure 8(c). As for χ″, the transition point where χ″ increases rapidly does not change while the absolute value of χ″ is suppressed by a static magnetic field. Since the cusp is from the spin freezing scenario in the SG system, the static magnetic field cannot change the overall collective behavior where most of the magnetic moments freeze but it can decrease the magnetic response of freezing magnetic moments under an oscillating field. These observations suggest that the cusp temperature is attributed to an SG behavior in KLaMn$_{1-x}$Te$_4$.

In order to further reveal the characteristics of the SG behavior, $T_f$ was extracted from the cusp position in each frequency in χ′, see Figure 9(a). A relative change of $T_f$ with the magnitude of frequency is calculated as the following formula:[37]

$$K = \frac{\Delta T_f}{T_f \log(\Delta f)}$$

This gives an estimated parameter $K$ of ~ 0.064, which is much larger than the canonical SG system such as Cu$_{1-x}$Mn$_x$ ($x$ ~ 1%) alloys ($K$ ~ 0.005) with a very dilute concentration of Mn magnetic atoms but below the value in the normal superparamagnets ($K$ ~ 0.1–0.3).[37, 49] A critical slowing down approach was also employed to study its dynamic properties of SG in KLaMn$_{1-x}$Te$_4$ with the following equation[37, 50]:

$$\tau = \tau_0 (\frac{T_f}{T_{SG}} - 1)^{-zv}$$

where $\tau$ is the inverse of frequency $f$, $\tau_0$ is the characteristic spin relaxation time, $T_{SG}$ is the spin freezing temperature when frequency goes to zero, $zv$ is the dynamic critical exponent. The fitting by the above equation gives $zv$ = 11.7, $\tau_0$ = 2.17 × 10$^{-7}$s and $T_{SG}$ = 4.48 K, see Figure 9(b). The value of $T_{SG}$ in a zero-frequency limit is consistent with the freezing temperature ($T_f \sim$ 5 K) in the DC magnetic susceptibility. $zv$ falls in the range between 4 and 13 for a SG system.[37] $\tau_0$ is much larger than the case in a canonical SG system which has a magnitude of $10^{-12} - 10^{-13}$ s, pointing to a rather slow spin dynamic relaxation process.[37] The above analysis supports the presence of cluster SG behavior in KLaMn$_{1-x}$Te$_4$ which is a natural consequence of its actual crystal structure and composition as explained below.

There are several ingredients that drive a system to SG behavior including randomness (disorder), competing interaction and magnetic frustration[37, 49], all of which are present in KLaMn$_{1-x}$Te$_4$. First, there are Mn vacancies randomly distributed in the [Mn$_{1-x}$Te] ($x$ = 0.35) layers. In fact, the ordered-Mn-vacancy compounds $A_2$Mn$_3$Se$_4$ ($A$ = Rb, Cs)[47] and BaFMn$_{0.5}$Te[20] show antiferromagnetic order. Second, massive amount of Mn vacancies ($x \sim$ 0.35) creating distributed Mn-Mn distances in the Mn sublattice, can lead to both antiferromagnetic and ferromagnetic interactions depending on the actual super-exchange path in the [Mn$_{1-x}$Te] layer. As a result, competing interactions are expected to exist in the [Mn$_{1-x}$Te] layers. In contrast to a canonical SG with diluted magnetic atoms such Cu$_{1-x}$Mn$_x$ alloys ($x \sim$ 1%)[37, 49], the Mn magnetic ions in KLaMn$_{1-x}$Te$_4$ are rather densely packed, making it possible to form magnetic clusters. A rather high spin relaxation time $\tau_0$ observed in KLaMn$_{1-x}$Te$_4$ points to the existence of magnetic clusters because it will take a much longer time to drive large non-equilibrated magnetic clusters to equilibrated ones than the case in a single isolated magnetic ion.

**Specific Heat.** No obvious anomaly at ~ 5 K is observed in the specific heat data of KLaMn$_{1-x}$Te$_4$, see Figure 10(a), further supporting that the cusp in the magnetic susceptibility is due to a SG behavior rather than a classical magnetic phase transition. The low temperature specific heat in KLaZn$_{1-x}$Te$_4$ can be well described by the formula $\gamma T + \beta T^3 + \delta T^5$ where $\gamma T$ is the electronic specific heat and $\beta T^3 + \delta T^5$ is the phonon specific heat at low temperatures,[51] see Figure 10(b). The obtained electronic coefficient $\gamma = 3.7$ mJ K$^{-2}$ mol-fu$^{-1}$ is larger than that in LaTe$_3$ (1.1 mJ K$^{-2}$ mol-fu$^{-1}$)[31] normalizing for the same number of LaTe$_3$ layers in the formula. This result is consistent with electron transfer from the [KZn$_{1-x}$Te] to [LaTe$_3$] layer and most likely to the Te-square net, creating a self-doping scenario in KLaZn$_{1-x}$Te$_4$. Since the main oxidation state of Mn element in KLaMn$_{1-x}$Te$_4$ is almost the same as Zn element in KLaZn$_{1-x}$Te$_4$, electron transfer from [KMn$_{1-x}$Te] to [LaTe$_3$] also exists. The estimated Debye temperature $\Theta = 181$ K from the equation $\Theta = (\frac{12NR\pi^4}{5\beta})^{1/3}$ is very close to LaTe$_3$[31] which is reasonable as they share the same building blocks and have a similar phonon spectrum.

Since KLaZn$_{1-x}$Te$_4$ and KLaMn$_{1-x}$Te$_4$ have the same crystal structure and the mass difference between Zn and Mn atoms is small, their phonon specific heat will be very similar. They have the same electron counts in the Te-square nets suggesting that they should also have similar electronic specific heat. Therefore, the total specific heat of KLaZn$_{1-x}$Te$_4$ will be a good approximation to the non-magnetic contribution in KLaMn$_{1-x}$Te$_4$ at low temperature. The magnetic specific heat was obtained by subtracting the specific heat of KLaZn$_{1-x}$Te$_4$ from that of KLaMn$_{1-x}$Te$_4$, see Figure 10(c). A broad peak in $C_m/T$ appears at ~ 20 K, much higher than $T_f$. This indicates that magnetic entropy loss happens way above $T_f$, which is a typical feature of a SG system.[37, 49] The integrated magnetic entropy is approaching the theoretical value of a spin-1 system $R\ln(3)$ with increasing temperature, see Figure 10(d). Therefore, the specific heat data confirm the metallic behavior in KLaZn$_{1-x}$Te$_4$ and supports a SG behavior in KLaMn$_{1-x}$Te$_4$.

## Conclusions

Two Q2D heterostructures KLa$M_{1-x}$Te$_4$ ($M$ = Mn, Zn) were successfully synthesized, composed of practically lattice-matched alternating [LaTe$_3$] and [K$M_{1-x}$Te] layers. Electron transfer from [K$M_{1-x}$Te] to [LaTe$_3$] layers exists creating a self-doping effect, and at the same time maybe serving to stabilize the structures because the charge transfer between the layers is driven by their different chemical potential levels and modulates the Fermi energy of the [LaTe$_3$] affecting the details of Fermi surface nesting and the **q**-vector. KLaMn$_{1-x}$Te$_4$ exhibits CDW order with **q** = 1/2**b**$^*$ + 0.3478**c**$^*$ at room temperature originating from electronic instability of Te-square nets, and cluster SG behavior with $T_f \approx 5$ K coming from disordered [Mn$_{1-x}$Te] layers. In-plane lattice match and chemical compatibility between two functional layers are important to stabilize the strong interlayer bonding hetero-layered structures and this feature can be a viable basis for targeting productive synthesis directions for hetero-structural compounds.

## ASSOCIATED CONTENT

**Supporting Information**.

This material is available free of charge via the Internet at http://pubs.acs.org.

Figure S1: PXRD experimental patterns of polycrystalline samples of KLaMn$_{1-x}$Te$_4$ and KLaZn$_{1-x}$Te$_4$ and their simulated patterns of KLaMn$_{1-x}$Te$_4$ and MnTe$_2$. Figure S2: PXRD pattern of KLaZn$_{1-x}$Te$_4$ sample exposed in air for two days and simulated pattern of Te element. Figure S3: one energy dispersive spectroscopy (EDS) data of KLaMn$_{1-x}$Te$_4$. Figure S4: Specific heat data of two KLaMn$_{1-}$

$_x$Te$_4$ and two KLaZn$_{1-x}$Te$_4$ samples. Figure S5: electric contacts created by Pt sputtering and silver paste for transport property measurements. Figure S6: magnetic susceptibility from another batch of KLaMn$_{1-x}$Te$_4$ polycrystalline sample. FigureS7: high angle annular dark field image and its corresponding EDS mapping of KLaZn$_{1-x}$Te$_4$ Table S1-5: detailed modulation structure parameters of KLaMn$_{1-x}$Te$_4$. Experimental details of charge transport property, energy-dispersive X-ray spectroscopy and scanning and transmission electron microscopy measurements. A cif file of the modulated structure.

# AUTHOR INFORMATION

**Corresponding Author**

*E-mail: m-kanatzidis@northwestern.edu

**Notes**

The authors declare no competing financial interest.

# Acknowledgements

This work was supported primarily by the U.S. Department of Energy, Office of Science, Basic Energy Sciences, Materials Sciences and Engineering. Use of the Center for Nanoscale Materials, an Office of Science user facility, at Argonne National Laboratory was supported by the U.S. Department of Energy, Office of Science, Office of Basic Energy Sciences, under Contract No. DE-AC02-06CH11357. TEM work was performed by use of the EPIC, Keck-II, and/or SPID facility of Northwestern University's NUANCE Center, which has received support from the Soft and Hybrid Nanotechnology Experimental (SHyNE) Resource (NSF ECCS-1542205); the MRSEC program (DMR-1720139) at the Materials Research Center; the International Institute for Nanotechnology (IIN); the Keck Foundation; and the State of Illinois, through the IIN. This work was supported in part by the National Science Foundation through the MRSEC program (NSF-DMR 1720139) at the Materials Research Center (V.P. Dravid, electron microscopy investigation). J.-K. Bao acknowledges the Alexander von Humboldt Foundation for financial support in Germany.

Table 1. Crystal data and structure refinement for KLaMn$_{1-x}$Te$_4$ at 293 K.

| | |
|---|---|
| Empirical formula | KLaMn$_{0.65}$Te$_4$ |
| Formula weight | 724.1 g mol$^{-1}$ |
| Temperature | 293 K |
| Wavelength | 0.71075 Å |
| Crystal system | orthorhombic |
| Space group | *Pmnm*(01/2γ)*s*00 |
| Unit cell dimensions | $a = 4.4815(3)$ Å, $\alpha = 90°$ <br> $b = 21.6649(16)$ Å, $\beta = 90°$ <br> $c = 4.5220(3)$ Å, $\gamma = 90°$ |
| **q**-vector(1) | 1/2**b**$^*$ + 0.3478(3)**c**$^*$ |
| Volume | 439.05(5) Å$^3$ |
| Z | 2 |
| Density (calculated) | 5.4774 g cm$^{-3}$ |
| Absorption coefficient | 19.157 mm$^{-1}$ |
| *F*(000) | 601 |
| Crystal size | $0.13 \times 0.10 \times 0.007$ mm$^3$ |
| $\theta$ range for data collection | 2.82 to 29.2° |
| Index ranges | $-6 \leq h \leq 6, -30 \leq k \leq 30, -6 \leq l \leq 6, -2 \leq m \leq 2$ |
| Reflections collected | 18531 (4014 main + 14517 satellites) |
| Independent reflections | 3087 (716 main + 2371 satellites) [$R_{\text{int}} = 0.0275$] |
| Completeness to $\theta = 29.2°$ | 99% |
| Refinement method | Full-matrix least-squares on $F^2$ |
| Data / constrains / restraints / parameters | 3087 / 0 / 0 / 89 |
| Goodness-of-fit on $F^2$ | 3.22 |
| Final R indices [$I>2\sigma(I)$] | $R_{\text{obs}} = 0.0535$, $wR_{\text{obs}} = 0.1317$ |
| R indices [all data] | $R_{\text{all}} = 0.0728$, $wR_{\text{all}} = 0.1341$ |
| Final R main indices [$I>2\sigma(I)$] | $R_{\text{obs}} = 0.0385$, $wR_{\text{obs}} = 0.0912$ |
| R main indices (all data) | $R_{\text{all}} = 0.0390$, $wR_{\text{all}} = 0.0912$ |
| Final *R* 1$^{\text{st}}$ order satellites [$I>2\sigma(I)$] | $R_{\text{obs}} = 0.1049$, $wR_{\text{obs}} = 0.2122$ |
| *R* 1$^{\text{st}}$ order satellites (all data) | $R_{\text{all}} = 0.1451$, $wR_{\text{all}} = 0.2146$ |
| Extinction coefficient | 0.2250(150) |
| $T_{\text{min}}$ and $T_{\text{max}}$ coefficients | 0.0949 and 0.8553 |
| Largest diff. peak and hole | 4.51 and −4.01 e·Å$^{-3}$ |

$R = \Sigma||F_o|-|F_c|| / \Sigma|F_o|$, $wR = \{\Sigma[w(|F_o|^2 - |F_c|^2)^2] / \Sigma[w(|F_o|^4)]\}^{1/2}$ and $w=1/(\sigma^2(I)+0.0004I^2)$

**Figures and Captions**

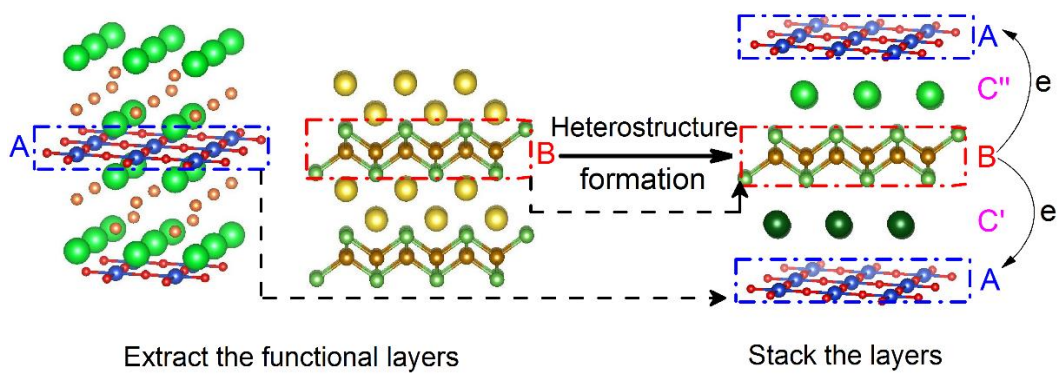

**Figure 1.** Schematic illustration of deriving heterolayered compounds. Certain crystal structure slabs A, B stable in parent compounds serve as the components in intergrowth compounds AB. The C layers show are generally charge-balancing counterions. Electron transfer between functional layers A and B can happen if they have different chemical potential levels, creating a self-doping effect.

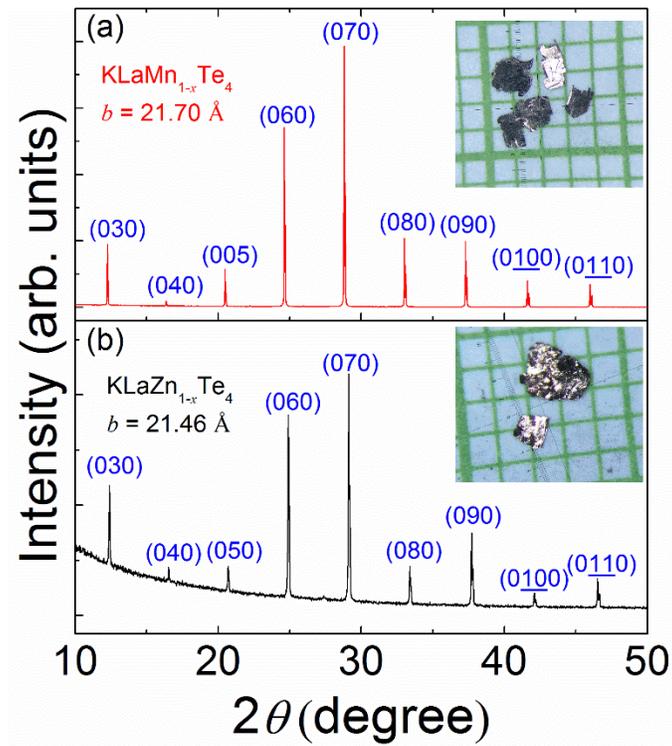

**Figure 2**. (a), (b) X-ray diffraction patterns of single crystals indexed by (0$k$0) reflections as well as pictures of single crystals under the optical microscope for KLaMn$_{1-x}$Te$_4$ and KLaZn$_{1-x}$Te$_4$, respectively.

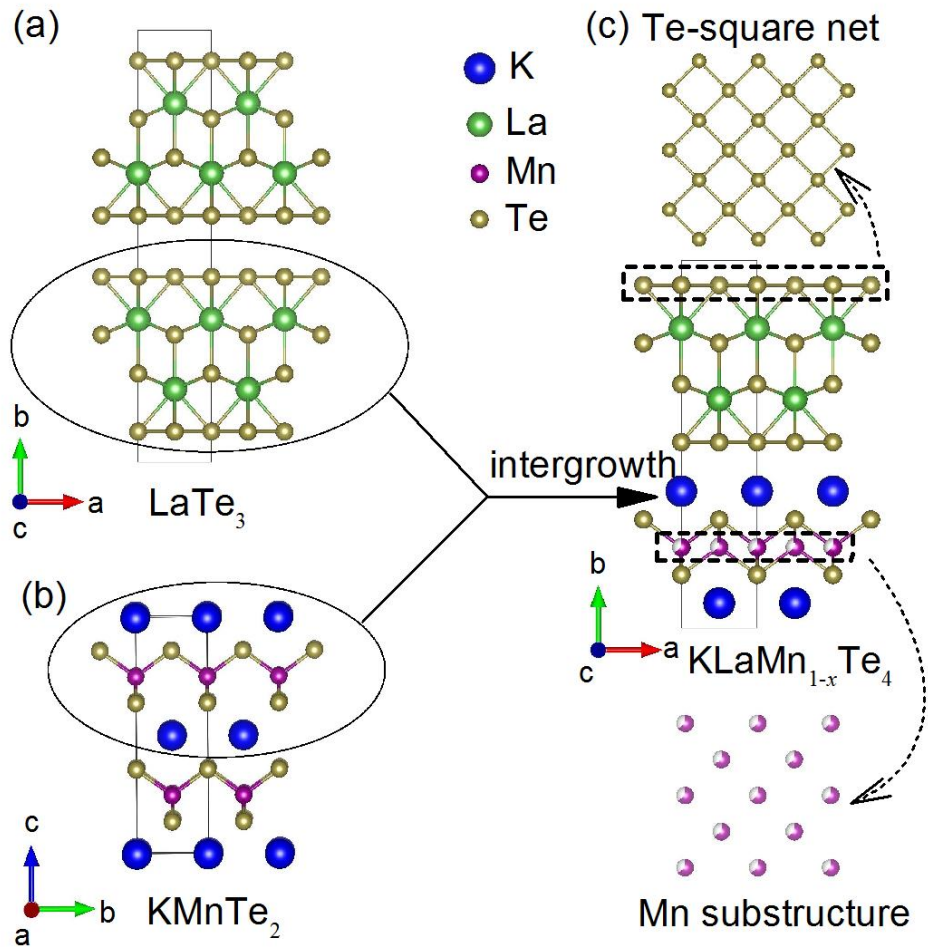

**Figure 3.** (a), (b) Crystal structures of LaTe$_3$ and KMnTe$_2$, respectively. (c) Average crystal structure of heterolayer compound KLaMn$_{1-x}$Te$_4$. Te-square nets and Mn substructure with vacancies shown in the upper and lower part of the panel, respectively.

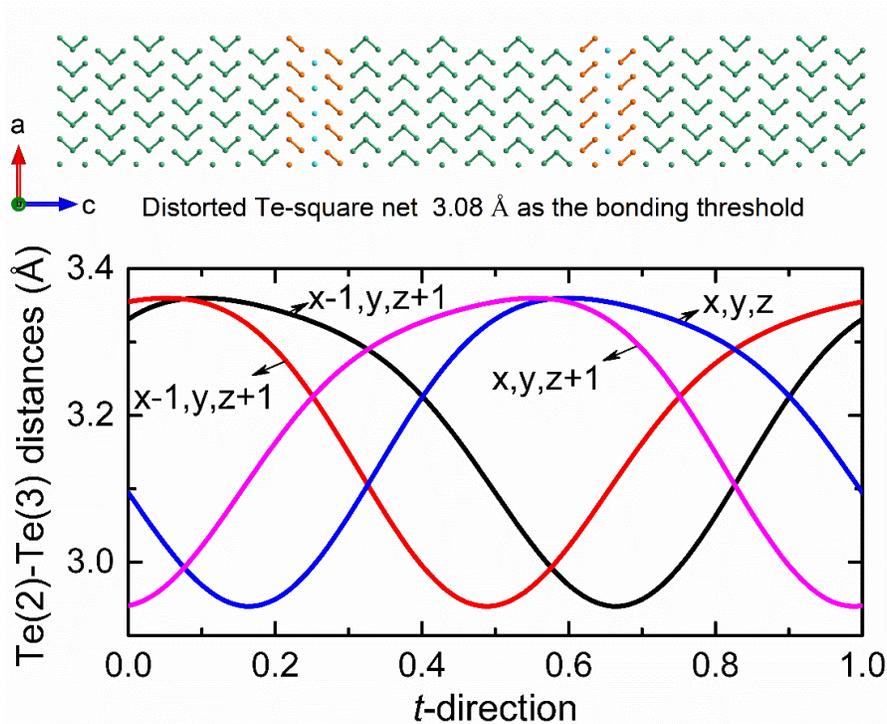

**Figure 4.** Upper panel: the pattern of monomers, dimers and trimers of Te atoms using the Te-Te bonding distance threshold of 3.08 Å in the distorted Te-square net. Lower panel: Te(2)-Te(3) distance with the modulation wave phase *t*. each Te(2) atom has four nearest Te(3) atoms.

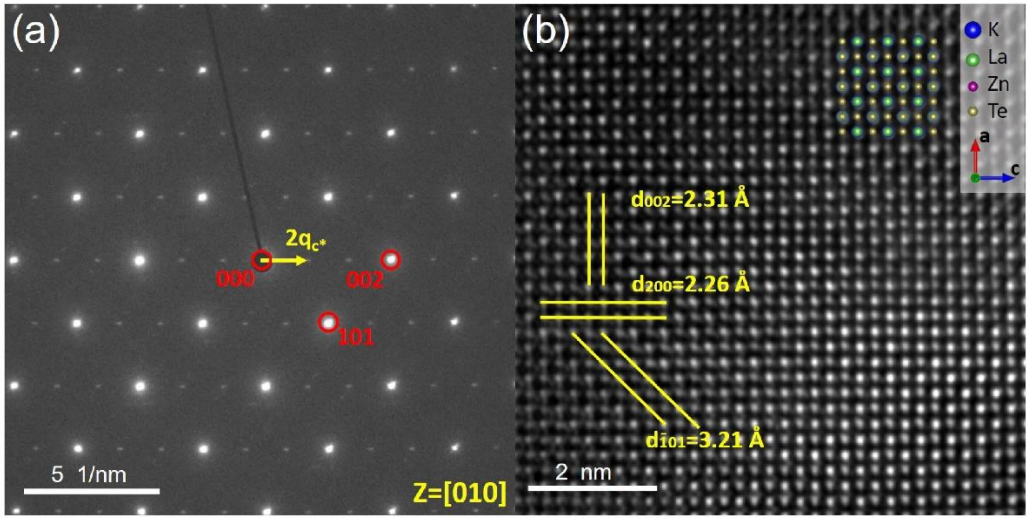

**Figure 5.** (a) Selected area electron diffraction (SAED) patterns of KLaZn$_{1-x}$Te$_4$ along the [010] zone axis. The indices of main reflections are based on the orthorhombic average structure (space group *Pmnm*). Satellite reflections with a second order are observed along **c**\* as labelled as 2**q**$_{c^*}$, indicating the existence of modulation structure. (b) HRTEM of KLaZn$_{1-x}$Te$_4$ along the [010] zone axis with some lattice planes and their corresponding distances marked. The structure schematic is overlaid on the upper right corner.

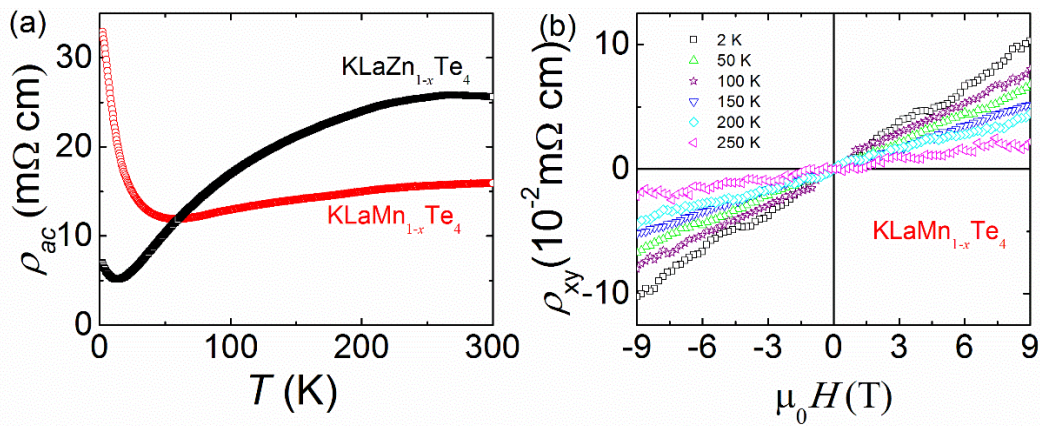

**Figure 6.** (a) Temperature dependence of resistivity in the *ac* plane for KLaMn$_{1-x}$Te$_4$ and KLaZn$_{1-x}$Te$_4$ single crystals. (b) Field-dependent Hall resistivity under different temperatures.

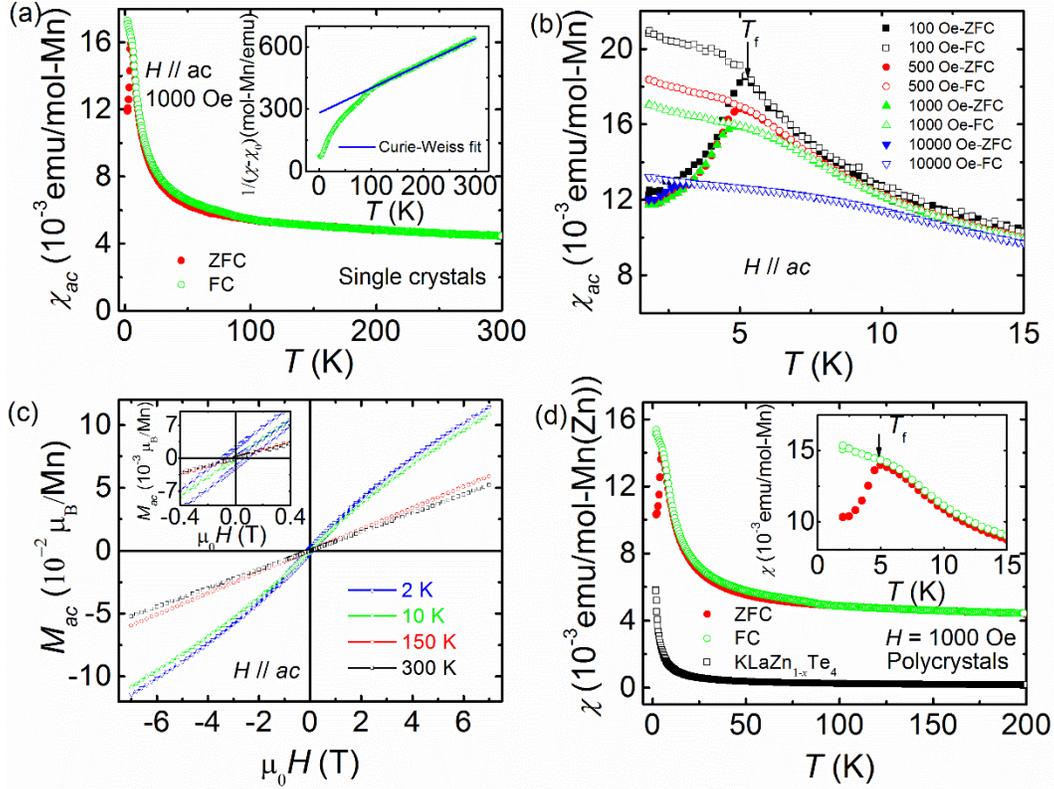

**Figure 7.** (a) Temperature-dependent magnetic susceptibilities of $KLaMn_{1-x}Te_4$ with applied fields (1000 Oe) in the *ac* plane under zero-field-cooled (ZFC) and field-cooled (FC) procedures. Inset shows the plot of $1/(\chi-\chi_0)$ vs $T$ for FC data with the blue dashed line as the Curie-Weiss fit above 150 K. (b) Temperature-dependent magnetic susceptibilities of $KLaMn_{1-x}Te_4$ under different magnetic fields. (c) Isothermal magnetizations versus fields at different temperatures. The inset shows the magnetizations under small magnetic fields. (d) Temperature-dependent magnetic susceptibilities of polycrystalline sample $KLaMn_{1-x}Te_4$ (circle symbol) under ZFC and FC procedures and $KLaZn_{1-x}Te_4$ (square symbol) under ZFC procedure. Inset shows the cusp feature below 15 K for $KLaMn_{1-x}Te_4$.

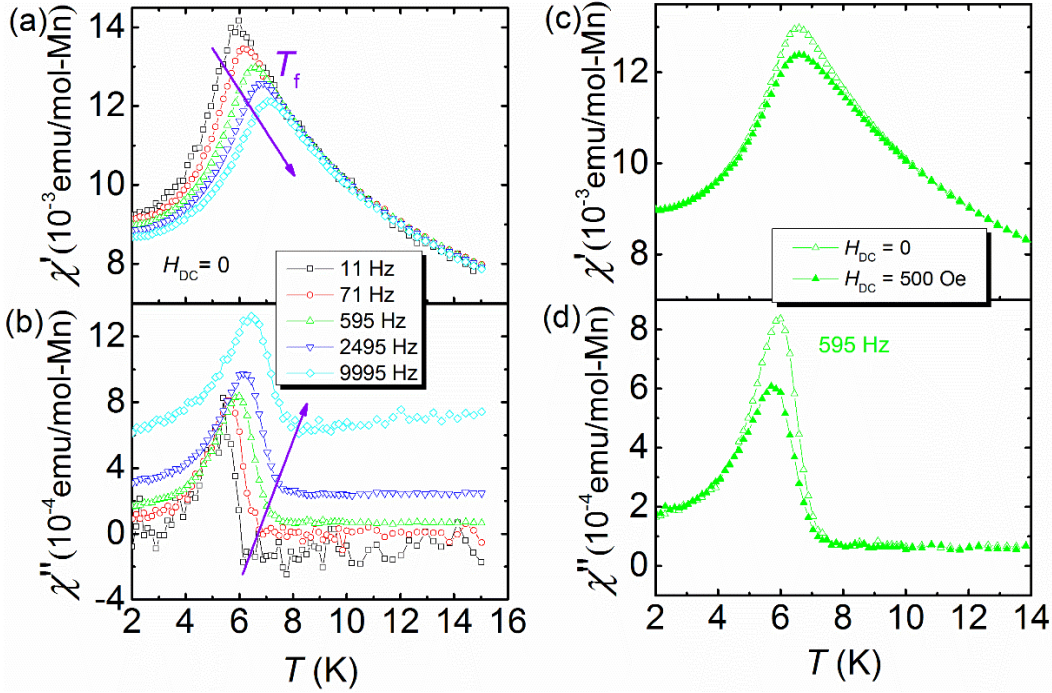

**Figure 8.** (a), (b) Real ($\chi'$) and imaginary ($\chi''$) components of temperature-dependent AC magnetic susceptibility with zero static magnetic field $H_{DC} = 0$ under different frequencies spanning from 11 to 9995 Hz. Some curves under certain frequencies are not shown to make the trend of the freezing temperature easy to identify. (c), (d) Temperature-dependent $\chi'$ and $\chi''$ with both $H_{DC} = 0$ and 500 Oe under the same frequency of 595 Hz.

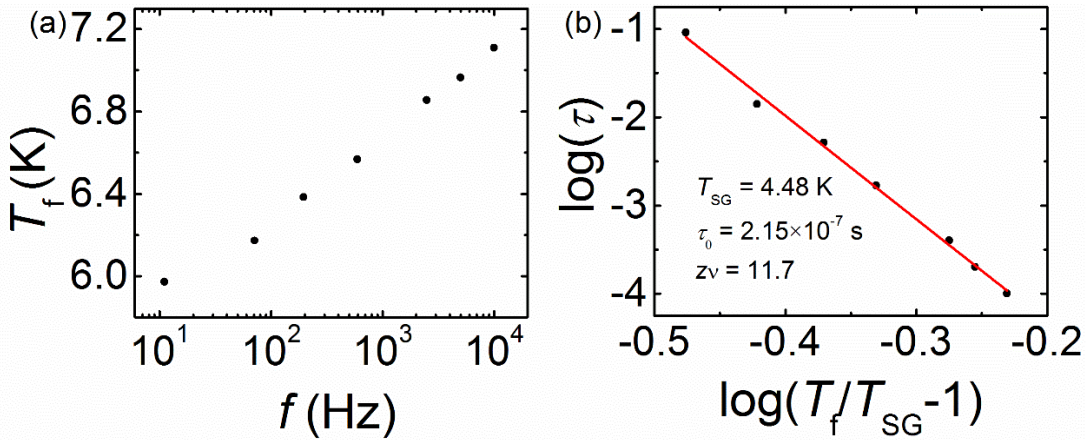

**Figure 9.** (a) Frequency-dependent freezing temperatures $T_f$ extracted from the cusp temperatures in the real ($\chi'$) component of AC magnetic susceptibility. The axis of frequency is plotted as a log scale. (b) Critical slowing down plot with the formula $\tau = \tau_0 (\frac{T_f}{T_{SG}} - 1)^{-zv}$. The red line is the fit for the experimental data.

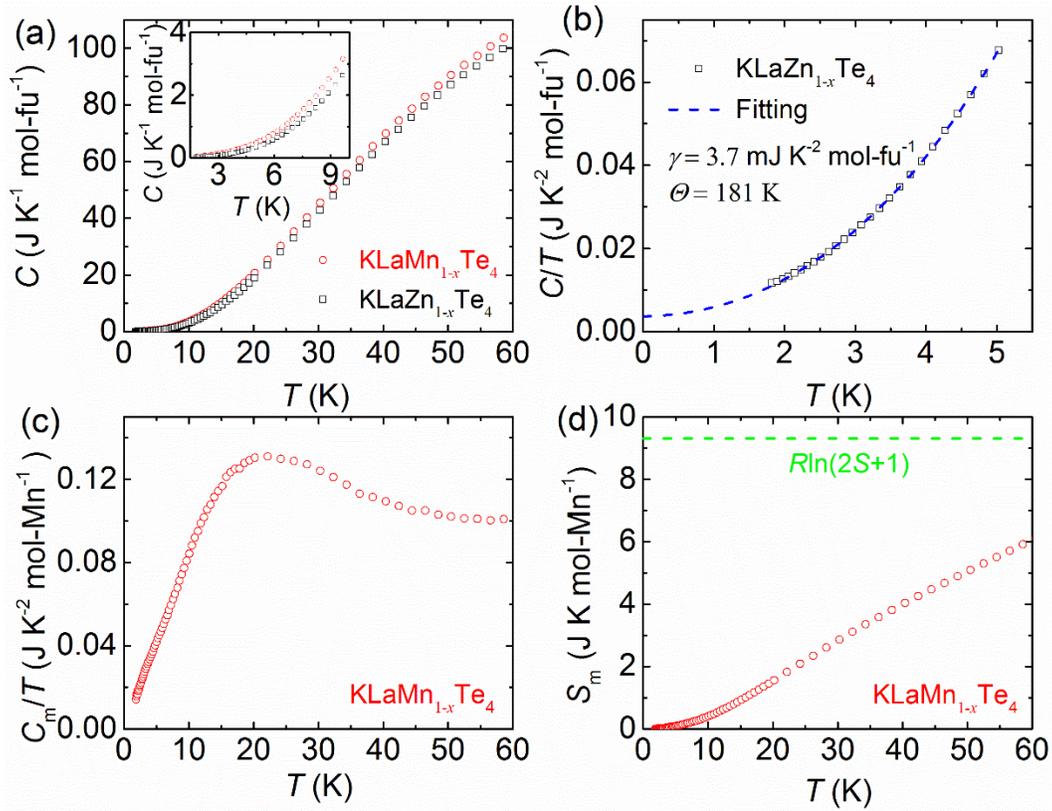

**Figure 10**. (a) Temperature-dependent specific heat $C$ for KLa$M_{1-x}$Te$_4$ ($M$ = Zn, Mn). Inset shows the data below 10 K. (b) Specific heat data fitting by considering phonon and electronic contribution for KLaZn$_{1-x}$Te$_4$. (c) Temperature dependence of magnetic contribution of specific heat over temperature $C_m/T$ after subtracting phonon and electronic contribution in KLaMn$_{1-x}$Te$_4$. (d) Magnetic entropy $S_m$ after integrating $C_m/T$ curve in (c). The green dashed line corresponds to the value of total magnetic entropy loss of a spin-1 $R\ln(3)$.

**For Table of Contents entry only**

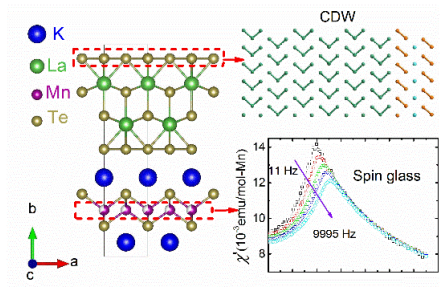